\documentclass{sig-alternate-05-2015}
\usepackage[utf8]{inputenc}
\usepackage{url}
\usepackage{dirtree}
\usepackage{csquotes}

\usepackage[font=small,labelfont=bf]{caption}

\usepackage{float}
\floatstyle{ruled}
\newfloat{listing}{h}{lop}
\floatname{listing}{Listing}

\usepackage{listings}
\usepackage{etoolbox}
\makeatletter
\patchcmd{\maketitle}{\@copyrightspace}{}{}{}
\makeatother

\usepackage{hyperref}
\hypersetup{
  pdfinfo={
    Author={Pascal Mainini, Annett Laube},
    Title={Access Control in Linked Data Using WebID},
    Producer={},
    Creator={}
  },
  pdfborder={0 0 0},
  colorlinks,
  citecolor=black,
  filecolor=black,
  linkcolor=black,
  urlcolor=black
}

\usepackage{xcolor}

\definecolor{light-gray}{gray}{0.95}

\begin{document}

\setcopyright{rightsretained}

\title{Access Control in Linked Data Using WebID}
\subtitle{A Practical Approach Validated in a Lifelong Learning Use Case}

\numberofauthors{2}
\author{
\alignauthor
Pascal Mainini\\
       \affaddr{Bern University of Applied Sciences (BFH)}\\
			 \affaddr{Institute for ICT Based Management}\\
       \affaddr{Höheweg 80, CH-2502 Biel/Bienne}\\
       \email{pascal.mainini@bfh.ch}
\alignauthor
Prof. Dr. Annett Laube-Rosenpflanzer\\
       \affaddr{Bern University of Applied Sciences (BFH)}\\
			 \affaddr{Institute for ICT Based Management}\\
       \affaddr{Höheweg 80, CH-2502 Biel/Bienne}\\
       \email{annett.laube@bfh.ch}
}
\date{20 April 2016}

\maketitle
\begin{abstract}
Linked Data technologies become increasingly important in many domains. Key factors for their breakthrough are \emph{security and trust}.
Classical means for access control lack granularity when \emph{parts} of the Linked Data graph must be protected. The WebID, combining 
semantic web concepts with methods from certificate based authentication and authorization, seems promising to fulfill all requirements 
concerning security and trust in the semantic web.

In the PerSemID project, we challenged the WebID technology with a \emph{fully implemented proof-of-concept (PoC)} addressing a workflow coming from 
the domain of lifelong learning and student mobility. In our use case of study enrollment, we used WebIDs for authentication and to grant access 
to parts of triple stores, during cross domain triple store interactions to exchange data between stakeholders.
\end{abstract}

\ccsdesc[500]{Security and privacy~Authentication}
\ccsdesc[500]{Security and privacy~Access control}
\ccsdesc[500]{Security and privacy~Authorization}
\ccsdesc[500]{Security and privacy~Usability in security and privacy}

\printccsdesc

\keywords{Semantic Web, WebID, Linked Data, Access Control, Authentication, Authorization}

\section{Introduction}
PerSemID\cite{persemid}, the successor of  the CV3.0 project\cite{cv3}, investigated 
issues remaining open in practical applications of the WebID\cite{webid} technology: its use for authentication and authorization in triples stores. 

While not questioning WebID's general security properties -- they are implied by the underlying
mechanisms based on client certificate authentication given by Transport Layer Security (TLS) -- we investigated
the question of trust, or more specifically the question of \emph{level of assurance (LoA)}\cite{iso29115} in WebIDs.
The LoA, an important concept in identity and access management,  states a quality level regarding authentications.

The second aspect concerns the application of WebID for access control to resources, operated by independent parties
and in distributed environments. Here, we focused on triple stores and platforms for document management.

Our use case in the domain of lifelong learning and student mobility shows the use of Linked Data for administrative processes
in enrollment for studies.
The concepts developed
in our complete proof-of-concept prototype can easily be adapted to similar processes in other domains.

\section{Related Work}
\label{sec:relwork}
PerSemID lies at the intersection of two domains: identity and access management (IAM) and semantic web based technologies with a focus on attribute transfer and document management.

Web Access Control\cite{wac} is one of the first approaches in providing authorization based on WebIDs acting on the level of HTTP resources. 

Universal Access Control (UAC)\cite{uac}, which we used in \\* CV3.0, goes further and provides access control at the level of individual triples.
Like UAC, the {Privacy Preference Ontology}\cite{ppo} provides access control at triple level.

WebID+ACO\cite{aco} is an ontology for authorization which focuses on adding a role-based authorization model to HTTP.

S4AC\cite{s4ac} is a vocabulary for creating access control policies focusing on named graphs. S4AC is used by the SHI3LD project\cite{shi3ld}
for specifying permissions.

The MyProfile project offered an IDP service for WebID as well as a platform for social networking.
Online resources of MyProfile are not available anymore, detailed information can be found in Sambra\cite{sambra}.
Recently, a new initiative called Solid\cite{solid} seems to take up on MyProfile.

\section{WebID}
\label{sec:webid}
\begin{figure}[ht]
\centering
\includegraphics[width=7cm]{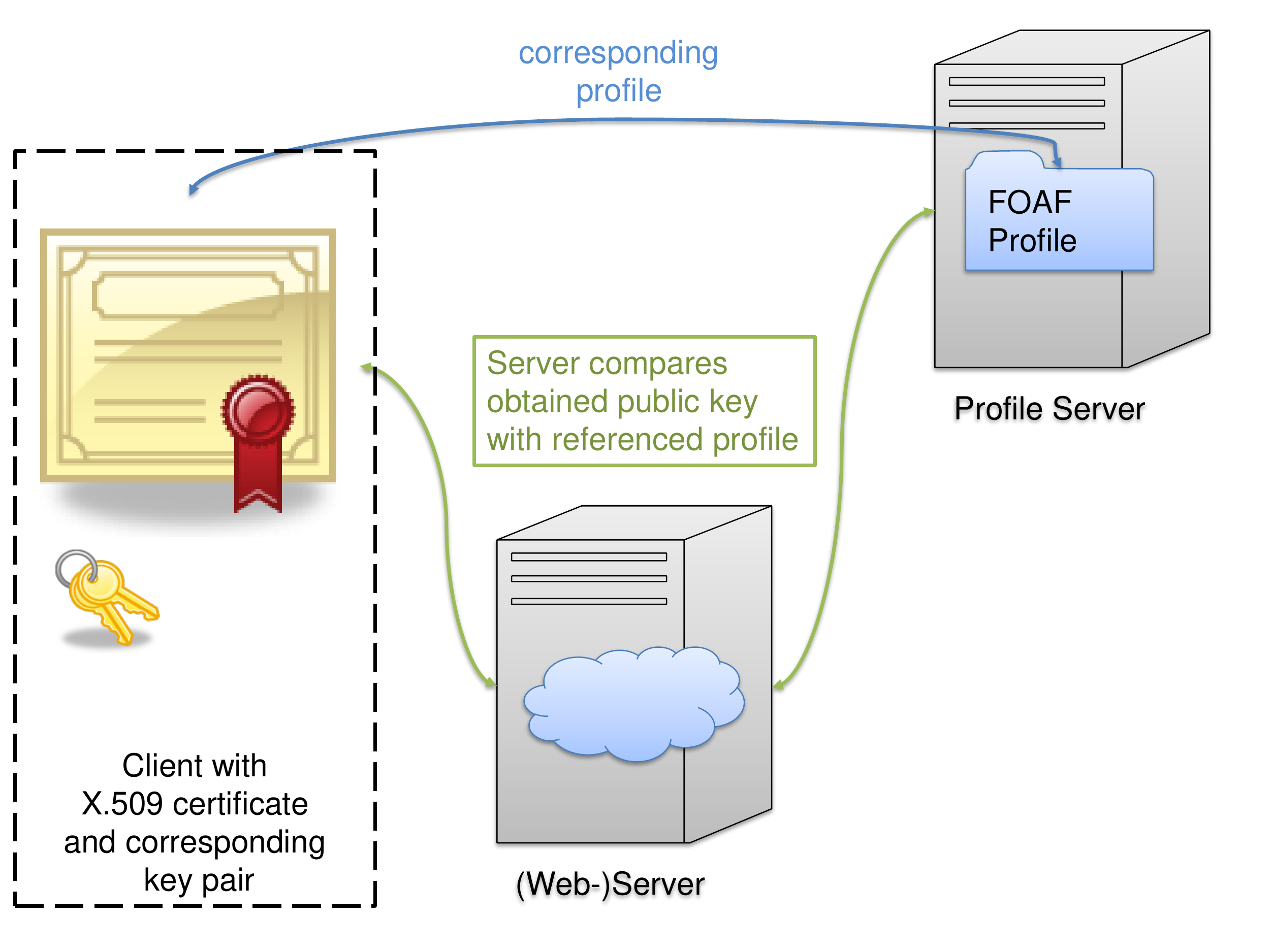}
\caption{WebID working principle}
\label{fig:webid}
\end{figure}

WebID authentication builds on the authentication of a client using X.509\cite{x509} client certificates. 
Functionality for using such certificates is present in all major browsers.
\\* To deliver additional information (for example personal attributes) and to establish URIs as identificator for a 
particular entity, WebID references a FOAF profile\cite{foaf} using a standard extension of X.509
(the \emph{Subject Alternate Name (SAN)} field). Figure~\ref{fig:webid} gives an overview of authentication using WebID.
The client (identified by its X.509 client certificate with corresponding key pair) wants to authenticate to
an application running on a (web-)server. The webserver retrieves the FOAF profile referenced in the SAN of the certificate and compares the information about the public key against the information obtained from the 
client certificate in the TLS handshake. If they match, authentication is successful. If desired, additional potentially signed attributes and other
information can be retrieved from the profile.

Anyone can issue a self-declared WebID by simply generating an appropriate certificate and publishing a corresponding
FOAF profile document on a webserver. 

The search for trust in WebID is a general problem which can also be found in other systems that use public key
cryptography -- as for instance classical X.509 certificates for websites or secure mail. Today, still the most common approach in
creating trust is the hierarchical Public Key Infrastructure (PKI) model with certificate authorities.

SuisseID, a PKI operated according to national signature laws\cite{zertes} by privately held certificate authorities 
(accredited by the state), provides X.509 certificates for authentication and digital signing on hardware tokens.
Besides the PKI, SuisseID also runs an attribute authority which provides additional information
about the holder of a certificate, like name, date of birth or gender. 

Being widely recognised, accepted, and having a very high level of trust, SuisseID would be an ideal partner for strengthening 
the LoA of a WebID. Furthermore, the attribute authority functionality could seamlessly be integrated into the FOAF profile server,
thus providing the same attributes with the same level of assurance for the Linked Data world.

Extending SuisseID with WebID is technically not a hard problem: Certificates issued by SuisseID must be extended to include
the proper SAN extension, containing the URI to the corresponding FOAF profile, and the issuing certificate authority must operate a webserver for serving these FOAF profile
documents accordingly. 

We took on the integration approach as described and validated it prototypically using the demo SuisseID identity provider from~\cite{suisseidsdk} (details are in~\cite{santomauro}).

Even though not being a new technology (surfaced end of 2008), WebID has not found broad adoption
so far.
While looking simple and flexible at first sight, it suffers from some issues which have been noted by 
others~\cite{foafprotocols} already. Most notably, the overall user experience of WebID 
seriously hinders broad adoption of the technology and unfortunately, there seems to be no intention on the part of
browser vendors to change this anytime soon. Additionally, nowadays the user typically owns a multitude of different devices, making certificate 
management nearly impossible.

\section{Use Case}
\label{sec:usecase}
Our PoC aimed to challenge the application of WebID in a working implementation of a real-life scenario.
The workflow conducted in the enrollment for master studies was chosen as our exemplary use case. This workflow involves
three primary actors: A \emph{student} who has successfully obtained a bachelor's degree (and may have additional qualifications), 
the institution at which this degree has been obtained (called \emph{bachelor university}) and finally the institution at which the
student whishes to enroll for master studies (the \emph{master university}).

A fully working prototype  has been made available~\cite{sourcecode}.
We also produced a screencast~\cite{screencast} demonstrating the main 
workflow between all involved parties 

From a technical perspective, PerSemID  builds upon the concepts of
a personal, semantic curriculum vitae developed in CV3.0. The architecture for a corresponding platform
for serving and maintaining such a CV has been defined in CV3.0's \emph{Content Access Service (CAS)}\cite{cv3cas}.
The CAS is a RDF triple store with additional document management capabilities as well as
an access control layer.

\subsection{Actors and Their Actions}
\label{sec:actors}
\begin{figure}
\centering
\includegraphics[width=7cm]{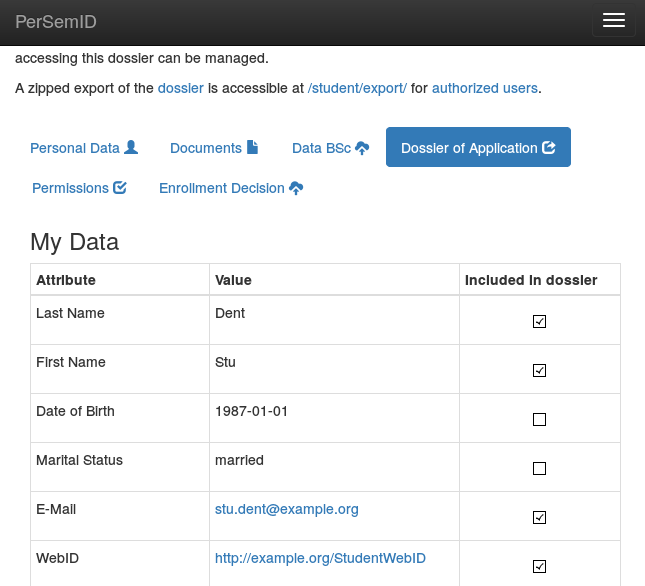}
\caption{Configuration of dossier of application by the student}
\label{fig:dossier}
\end{figure}

In a first step, the student prepares a \emph{dossier of application} which contains all relevant information
about the degree obtained and possible additional data in form of documents. Provenance of this data is either personal
information entered by the student directly or data obtained from the bachelor university in the \emph{bachelor dossier}.
The bachelor dossier is issued by the bachelor university as a single file, containing Linked Data about the degree obtained 
and optionally additional documents.
All this data is then stored in the student's CAS and the student can freely 
choose to include/exclude data per application at a master university (see Figure~\ref{fig:dossier}).

After having created the dossier of application, the student authorizes the master university to access the dossier 
by restricting access based on the university's WebID which is assumed to be publicly available.

Next, the addressed master university picks up the dossier by accessing it on the student's CAS.
Following a review of all the material in the dossier, a decision regarding acceptance to master studies can be made.
Now, the master university in turn stores its decision in its CAS and authorizes the WebID, given by the student, to access it.

In the last step, the student finally retrieves the decision from the master university.

\pagebreak

\subsection{Architecture and Implementation}

\subsubsection{Content Access Service}
There is no ready-made product similar to a content access service as specified by \cite{cv3cas}, thus the needed 
functionality had to be implemented in the PoC itself. A large range of (mature) triple stores is available (see in \cite{triplestores}). We chose to use Apache Jena\cite{jena}, that supports SPARQL 1.1 update together with other requirements.

A deliberately reduced set of document management capabilities has been implemented in the PoC code itself.

The {CAS} serves as storage for all metadata related to each actor and also as location for all application-specific configuration data, like
file system paths or granted permissions.
An example for the contents of the student's graph, including a granted permission for the WebID \texttt{hmsc.example.org} can be seen in Listing~\ref{listing:student}.

\begin{listing}
\begin{lstlisting}[emph={permission},emphstyle=\underbar]
@base <http://example.org/Student> .
@prefix rdfs: <http://www.w3.org/2000/01/rdf-schema#> .
@prefix xsd: <http://www.w3.org/2001/XMLSchema#> .
@prefix s: <http://persemid.bfh.ch/vocab/student#> .

<#> a s:Student ;
    s:webid <http://example.org/StudentWebID> ;
    s:name "Dent" ;
    s:vorname "Stu" ;
    s:email "stu.dent@example.org" ;
    s:matrikelnummer "1-234-56" ;
    s:permission <http://hmsc.example.org/webid#id> .
\end{lstlisting}
\caption{Partial example data of a student}
\label{listing:student}
\end{listing}

Documents, which can be uploaded by the student and the bachelor university, are given a unique ID and stored on the file system.
Metadata needed by the server for interacting with them is again stored in the triple store, in the named graph of the respective actor.

\subsubsection{Server Application}
The whole server application has been written in JavaScript and is running on node.js\cite{node}. HTTP-functionality has
been realized using the widely deployed middleware layer connect\cite{connect} which makes creation of applications serving
a variety of different requests straightforward. All communication between the frontend application and the server runs 
over a single HTTPS port.

\begin{figure}
\centering
\includegraphics[height=4.5cm]{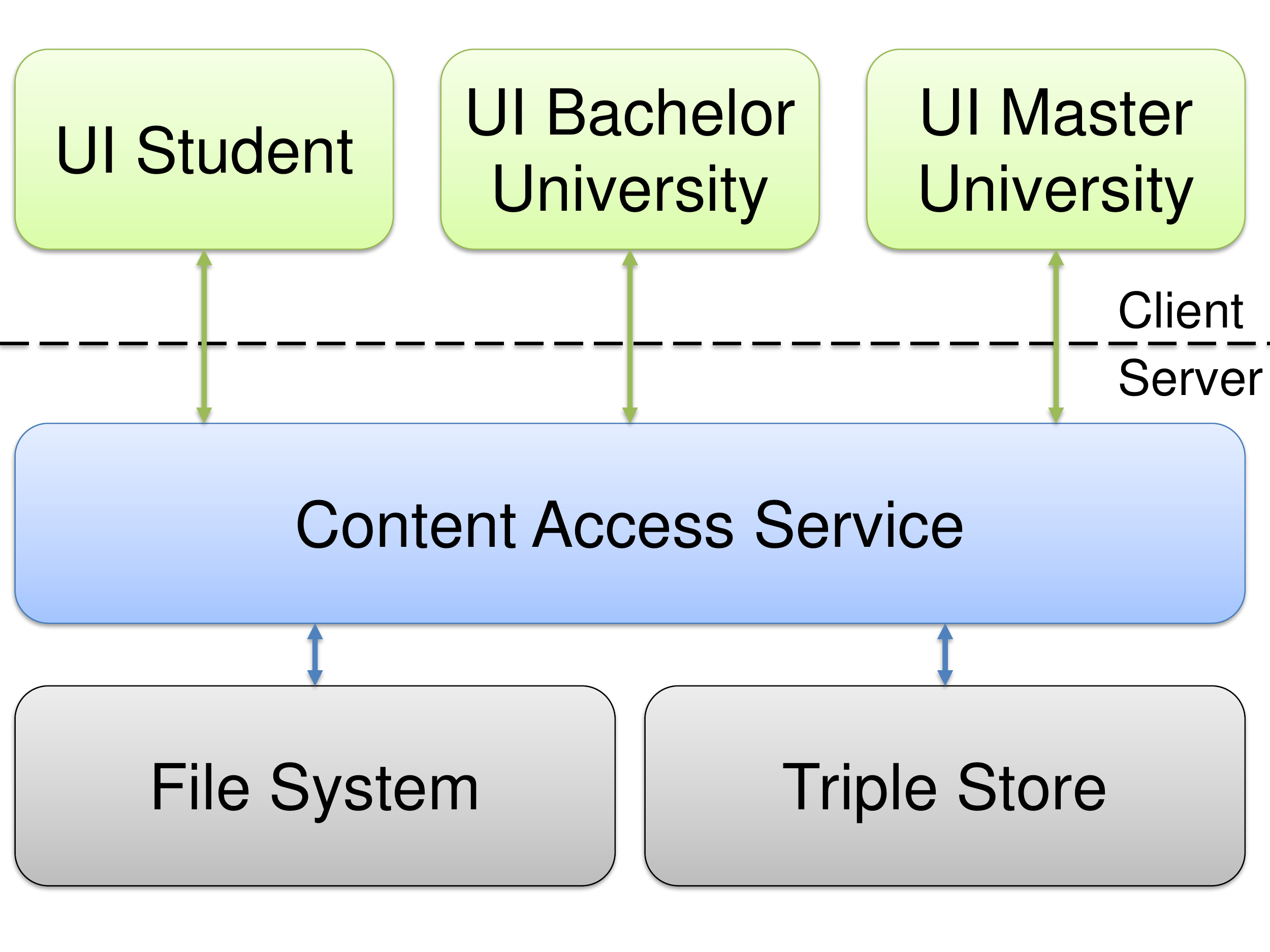}
\caption{Overview of architecture}
\label{fig:architecture}
\end{figure}

On both sides, client and server, we make use of  JavaScript RDF libraries like rdf-ext\cite{rdfext} 
and ld2h\cite{ld2h}. An overview of the architecture is given in Figure~\ref{fig:architecture}.

\subsubsection{WebID Identity Provider}
\label{sec:webidp}
All functionality needed for WebID authentication has also directly been implemented in the PoC itself, based on our experience in
the implementation of the WebIDP\cite{webidp} application, an identity provider for WebID developed in CV3.0.
A dedicated URL of the webserver serves the FOAF profiles referenced by the client certificates.
The certificates for all actors were generated directly using OpenSSL with respective configuration files.

\subsubsection{Cross-Domain Triple Store Interaction}
\label{sec:zip}

As described in Section~\ref{sec:actors}, all actors follow a defined scheme of interaction. In this scheme, there are three data
exchanges: download of bachelor data by the student from the bachelor university, download of data from the student by 
the master university and finally, download of data from the master university by the student.

This can be generalized as a concept for sharing data between triplestores or \emph{cross-domain triple store interaction}.
Multiple methods for implementing such interactions could be thought of, we considered the following three: (1) cross-site sharing using HTTP access control, known as \emph{CORS}\cite{cors}; (2) proxying of data on the server side; and (3) explicitly channeling data through the client.

Being limited by the same-origin policy, that restricts how a document or script loaded from one origin can interact with a resource from another origin, a direct interaction between the client-side program logic and the content access 
service of the remote party in an exchange cannot be implemented -- even considering the fact, that in our PoC scenario, all content was 
served from the same server.  This problem could be circumvented with HTTP access control (CORS), which allows for a relaxation of the restrictions imposed on the client.
By doing so, we would face another problem: in order to be able to dynamically adjust the needed HTTP headers, parties exchanging data
would have to know each other in advance -- rather an unlikely situation in a real world scenario.

One notices, that this is shifting control towards the server, leading to another option where the server acts as a proxy for the data to be exchanged.
Being a seemingly straightforward approach, this method has some serious drawbacks as well. We would have strong concerns regarding
security if the server could be instructed by the client to act as an open proxy interacting with unknown destinations.
Also, for the purpose of our PoC, hiding the exchanges between actors is not optimal for the demonstration of the implemented functionality.

So we finally set with the third option and implemented a very explicit data exchange using ZIP-files which are downloaded by an actor from
the remote party and manually imported into their own CAS. While this may look odd or even ancient at a first glance, it has some great benefits
for our validation work, which amongst others are: (a) explicit WebID authentication and authorization are possible -- our main objective in this case; (b) separation of the actors and adminstrative borders are clearly visible; and (c) interaction with files is well known to the user.

\section{Conclusions}
Our prototype showing the process of study enrollment demonstrated, that by using Linked Data technologies, concrete and practical administrative workflows can be implemented easily and without hassles. Authentication and authorization using WebID stands the test regarding security requirements
in that area -- an integration into other, trusted identification systems such as SuisseID is technically possible and would enhance the WebID in terms of trust.

The prototype gave us insights in cross-domain triple store interaction and provided a model for future implementations of processes and workflows
based on Linked Data technologies. During the implementation, we encountered some issues, most notably related to the same origin policy of modern browsers.
For these issues, we gave an overview of possible solutions and described the one chosen.

Besides technical problems, our research clearly showed weak points in WebID.
Regarding broader acceptance of the technology, as means for authentication and especially as a \enquote{token} for permission handling, 
future work for better integration, portability and especially userfriendlyness must be undertaken. 
Here, we are particularly interested in approaches taken by recent projects like Solid -- and whether those will be successful in solving these issues.

\bibliographystyle{unsrt}
\bibliography{access-control-in-linked-data-using-webid}

\end{document}